\documentstyle[aps,epsf,epsfig,prl,psfig]{revtex}

\newcommand{\be}{\begin{equation}\FL}
\newcommand{\ee}{\end{equation}}
\newcommand{\beas}{\begin{eqnarray*}}
\newcommand{\eeas}{\end{eqnarray*}}
\newcommand{\bea}{\begin{eqnarray}\FL}
\newcommand{\eea}{\end{eqnarray}}

\newcommand{\eps}{\epsilon}

\begin{document}

\twocolumn[\hsize\textwidth\columnwidth\hsize\csname
@twocolumnfalse\endcsname

\title{Stylized facts of financial markets and
market crashes in Minority Games}
\author{Damien Challet${}^{(1)}$, Matteo Marsili${}^{(2)}$ and 
Yi-Cheng Zhang${}^{(3)}$} 
\address{${}^{(1)}$ Theoretical Physics, 
Oxford University, 1 Keble Road, Oxford OX1 3NP, United Kingdom\\
${}^{(2)}$ Istituto Nazionale per la Fisica della Materia ({\it INFM}),
Unit{\'a} di Trieste-SISSA, I-34014 Trieste, Italy\\${}^{(3)}$
Institut de Physique Th{\'e}orique, Universit{\'e} de Fribourg, 
Perolles CH-1700, Switzerland}
\date{\today} 
\maketitle 
\widetext
\begin{abstract}
We present and study a Minority Game based model of a financial market
where adaptive agents -- the speculators -- interact with
deterministic agents -- called {\em producers}. Speculators trade only
if they detect predictable patterns which grant them a positive
gain. Indeed the average number of active speculators grows with the
amount of information that producers inject into the
market. Transitions between equilibrium and out of equilibrium
behavior are observed when the relative number of speculators to the
complexity of information or to the number of producers are
changed. When the system is out of equilibrium, stylized facts arise,
such as fat tailed distribution of returns and volatility clustering.
Without speculators, the price follows a random walk; this implies that stylized facts
arise because of the presence of speculators. Furthermore, if speculators
abandon price taking behavior, stylized facts disappear.
\end{abstract}

\pacs{PACS numbers: 02.50.Le, 05.40.+j, 64.60.Ak, 89.90.+n}
\narrowtext

]

\section{Introduction}

Physicists' interest for Economy and more specifically financial
markets has exploded in last two years. An important part of the
research is the analysis of financial data
\cite{Stanleyret,Stanleyvol,MantegnaStanley}, which has led to the characterization of 
some empirical statistical regularities, known as ``stylized facts''.
Consequently a lot of work have attempted to build models of markets
that reproduce these properties in order to understand their cause. A
very promising way is to consider agents based models
\cite{Lux,Caldarellietal}. However, in spite of interesting
results obtained so far by numerical simulations, most of these models
are too complex and not well suitable to an analytic approach that  could explain
the origin of their complex behavior.

A different strand of literature, originated by the introduction of
the Minority Game (MG) \cite{CZ97}, has instead focused on highly
simplified toy models of financial markets\footnote{See \cite{web} for
a large collection of commented references about the MG.}. On one
hand, variants of this model have been shown to reproduce quite
accurately the stylized facts of financial
markets\cite{J99,J00,CCMZ,BouchMG}. On the other, there are analytic
approaches for this model which provide exact results for the limit of
infinitely many agents\cite{CMZe,MCZ,Coolen,MC,MMM}. These approaches
give a coherent picture of the collective properties and allow one to
investigate in detail a number of issues on the behavior of complex
systems of interacting agents, such as the role of market impact
\cite{MCZ,MC} and the interplay between different types of agents in a
market \cite{MMM}.

Our aim is to present what we believe to be the simplest MG giving
rise to a quite complex and rich behavior. In different regions of its
phase space, the model describes markets with gaussian statistics and
short ranged correlation, markets with fat tailed retuns and long
range correlations and even market crashes. In addition, it is
possible to obtain exact analytical results \cite{CMRZ} along the
lines of refs.  \cite{CMZe,MCZ}.

The first key ingredient of the model is the interplay between two
types of traders. The first type, called {\em producers} in
ref. \cite{ZMEM,MMM}, are traders who use the market for exchanging
goods; Their trading decisions originate from outside opportunities
related to the economic activity and not on the market dynamics
itself. These traders have a predictable behavior with respect to a
news arrival process and hence they inject information into the
market. They represent the underlying economic activity, so they
could also be called ``fundamentalists'': in the absence of other
types of agents market prices would follow a random walk, which we
could call the {\em fundamentals}.

The other type of traders are speculators. They are adaptive agents
with bounded rationality. They study the relationship between the news
arrivals and market reactions in order to anticipate market
movements. Their aim is to gain from market fluctuations. On one hand
they provide liquidity for producers, on the other they color the
white noise process produced by the latter and by the news arrival
process. Therefore they are responsible for the emergence of stylized
facts.

The second key ingredient of the model is that we allow the
speculators for the possibility of not trading if the market does not
contain sufficiently profitable arbitrage opportunities for them. In
the language of Physics, this makes the model grand canonical. Such
models have been studied in refs. \cite{J99,J00} and more recently in
ref. \cite{BouchMG}, where a fundamental mechanism for long-range
correlation of the volatility was proposed. However, the proposed
grand canonical mechanism in refs \cite{J99,BouchMG} does not account
for the risky nature of markets, by contrast to that we propose here, whereas
the one found in ref \cite{J00} is in essence the one we consider here
\footnote{Note however that it is not motivated by the risky nature of markets.
In particular, we do not have to introduce utility functions.}.

These two features -- the interplay between producers and speculators
and the possibility of not trading for speculators -- have been
introduced in ref. \cite{MMM}, though their consequences have not been
fully exploted. In addition, the grand canonical mechanism proposed
here is different and leads to qualitatively different results.

\section{The model}

We consider a set of agents who interact repeatedly in a market. In
each period $t=1,2,\ldots$, each agent $i$ chooses an {\em action}
$a_i(t)$, which is a real number. To fix ideas one may think that
$a_i(t)>0$ means that agent $i$ wants to buy $a_i(t)$\$ of an asset
whereas $a_i(t)<0$ implies that he wants to sell. 

\subsection{Market mechanism and information}

Following
ref. \cite{CCMZ}, we leave details at this level unspecified, and
directly define the excess demand $A(t)=\sum_{i=1}^N a_i(t)$ and 
the payoffs of agents:
\be
g_i(t)=-a_i A(t)
\ee
These payoffs are such that those agents who are in ``minority''
($a_iA(t)<0$) are rewarded. This captures the fact that when there
are a lot of buyers, sellers may sell at a higher price.

Following Refs. \cite{J99,Caldarellietal,Farmer,bouchaudcont}, we define a price
dynamics in terms of the excess demand, as
\be
\log p(t+1)=\log p(t)+r(t)=\log p(t)+\frac{A(t)}{\lambda}
\ee
where $r(t)$ is the return at time $t$ and $\lambda$ is related to the
market depth.
The market is also characterized by a {\em news arrival process} which
is modeled by an integer $\mu(t)$ which is drawn randomly\cite{cavagna,CM00} and 
independently in each period from the integers $1,\ldots,P$.
$\mu(t)$ labels the ``state of the world'' which encodes all relevant
economic information. 

\subsection{Producers}

The first type of traders -- the {\em producers} -- behave in a
deterministic way with respect to $\mu(t)$. This means that if $i$ is
a producer, $a_i(t)$ is a function of $\mu(t)$ only, i.e.
$a_i(t)=\sigma_i^{\mu(t)}$. For each agent $i$ and state $\mu$ we draw
randomly $\sigma_i^\mu$ from a fixed distribution. We shall take the
bimodal distribution $\sigma_i^\mu=\pm 1$ with equal probability, in
what follows\footnote{Any distribution with zero average and unit
variance leads to exactly the same results in the limit $N\to\infty$.}

Let $N_p$ be the number of producers and we introduce the reduced
number $n_p=N_p/P$ for convenience.

\subsection{Speculators}

The second type of traders -- the {\em speculators} -- are adaptive.
They are assigned a number $S+1$ of {\em trading strategies}: When
speculator $i$ uses strategy $s=0,1,\ldots,S$, then his action is
$a_i(t)=\sigma_{s,i}^{\mu(t)}$. For strategies $s>0$ -- called {\em active} --, 
$\sigma_{s,i}^\mu$ is again drawn
randomly and independently for each $s,i$ and $\mu$ from the bimodal 
distribution. For $s=0$, instead, $\sigma_{0,i}^\mu=0$ for all $\mu$.

In other words, producers can be regarded as speculators with just one
active trading strategy and speculators can decide not to trade if they
resort to their $0$-strategy. At odd with producers, speculators
have an additional degree of freedom, which is the choice of the
strategy $s_i(t)$ they will play at time $t$. In order to take this
decision, each speculator $i$ keeps track of the performance of each
of his trading strategy $s$ by assigning a {\em score} $U_{i,s}$ to
it.  The strategy $s_i(t)$ which agent $i$ follows at time $t$, is
that with the highest score. His action will then be
$a_i(t)=\sigma_{i,s_i(t)}^{\mu(t)}$. The scores are updated according
to
\be
U_{i,s}(t+1)=U_{i,s}-a_{i,s}^{\mu(t)}A(t)+\epsilon\delta_{s_i(t),0}.
\ee

This means that active strategies $s>0$ are ``rewarded'' by the
(virtual) gain they would have given to agent $i$ if they had been
played\footnote{This is true only if $A(t)$ would not have changed if
$i$ had actually played strategy $s$ instead of $s_i(t)$ (which is why
we speak of virtual gain). In other words, we are assuming that agents
neglect their market impact and behave as price takers. This is by no
means an innocent assumption as shown in refs. \cite{CMZe,MCZ,MC} (see
later).}. The $s=0$ strategy is instead rewarded by a constant amount
$\epsilon>0$. This implies that an agent is willing to use a trading
strategy only if it gives an average gain larger than $\eps$.  One can
interpret $\eps>0$ as modeling either a risk-free asset which ensures
a constant gain -- the bank interest rate -- or more simply a
risk-premium for not trading. The latter may reflects agent's risk
aversion to trading in a risky market. Note that with $\epsilon<0$ and
large enough, one recovers the standard MG because all agents use
active strategies.

There are $N_s$ speculators and we find it convenient to introduce the
reduced number $n_s=N_s/P$. Note that $n_s=1/\alpha$ where $\alpha$ is the
control parameter used in refs. \cite{Savit,CMZe,MCZ}.

\section{Results}

We performed extensive numerical simulations of the market model.  For
$P\gg 1$ all (intensive) quantities in the stationary state only depend on
the parameters $n_s$, $n_p$ \cite{Savit,CM,CMZe,MMM} and
$\epsilon$. All simulations were performed with $S=2$. As usual no
qualitative change is expected for $S>2$ \cite{MCZ}.

Before passing to the discussion of results, let us mention that the
model can be solved along the lines of ref. \cite{MCZ}. An account of
this solution will be presented in a forthcoming publication
\cite{CMRZ}.

\subsection{Market's ecology}

\begin{figure}
\centerline{\psfig{file=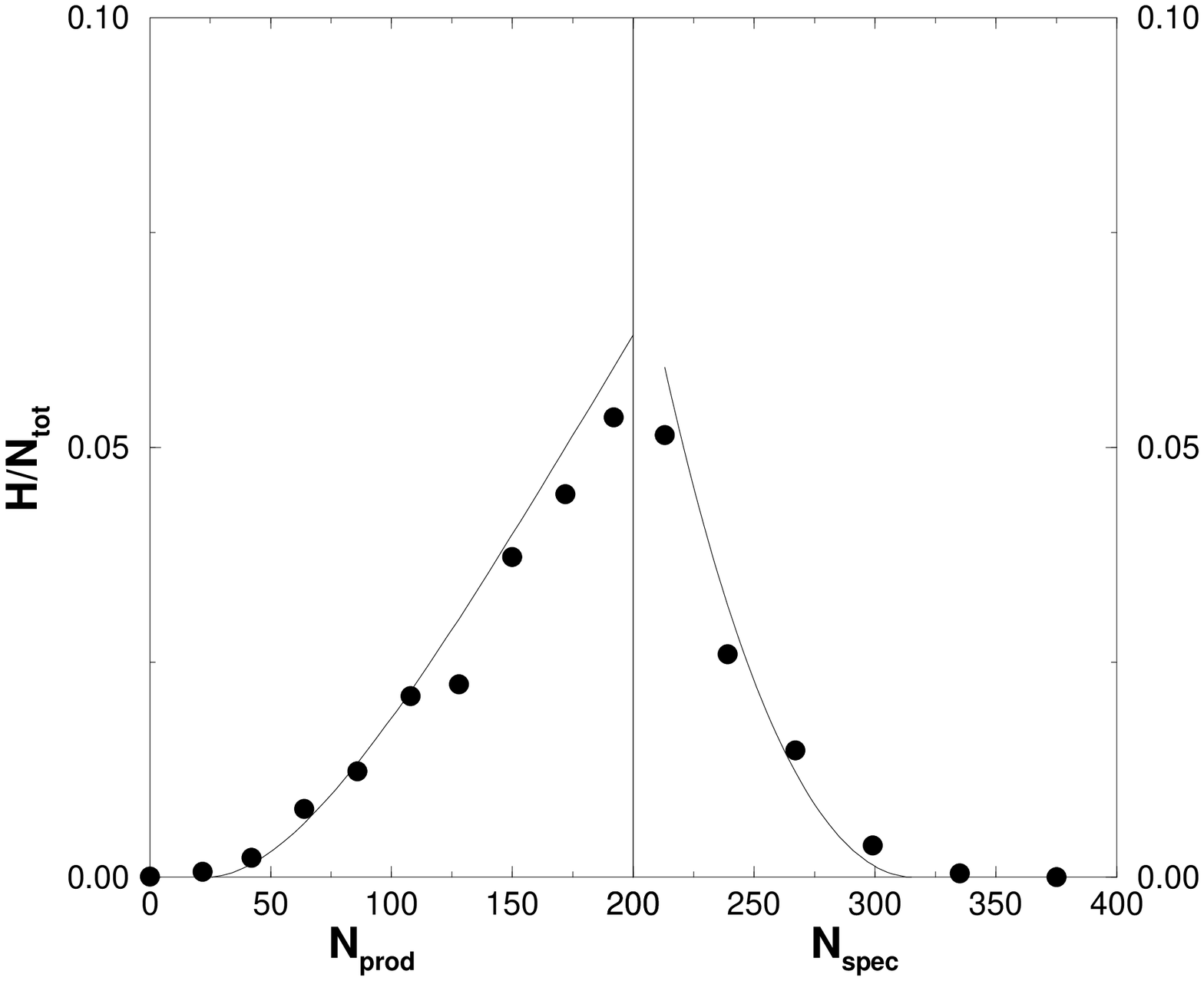,width=6cm}}
\caption{$N_p$ Producers are added to a standard MG ($P=64$, $N=200$,
$S=2$) where the outcome is unpredictable ($H=0$): information
content, or negative entropy, increases with $N_p$. Then, while $N_p$
is kept fixed at 200, additional speculators are added; they finally remove
completely the information put by producers (average over 100 realizations). The
continuous lines are analytical results from the exact solution.}
\label{hHh}
\end{figure}

The interplay between the two types of agents is shown in Fig. \ref{hHh}
for $\epsilon=-\infty$ (no $0$-strategies): keeping the number of speculators
fixed and increasing the number of producers, the market predictability
\be
H=\frac{1}{P}\sum_{\mu=1}^P\langle A|\mu\rangle^2
\ee
increases\footnote{Here $\langle A|\mu\rangle$ is the average of
$A(t)$ conditional on $\mu(t)=\mu$. If this quantity is non-zero, then
the market is statistically predictable. $H$ measures predictability
and is related to negative entropy.} (left panel). When speculators
are added again to the market (right panel) they exploit
predictability and hence reduce it.

\begin{figure}
\centerline{\psfig{file=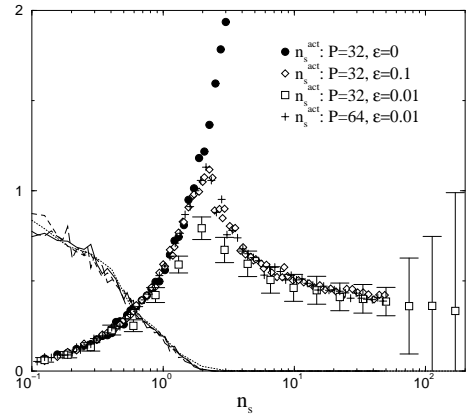,width=6cm}}
\caption{Predictability per agent $H/(N_p+N_s)$ (lines) and number of
active speculators $n_s^{\rm act}$ (symbols) as a function of $n_s$
for $n_p=1$ and some values of $\epsilon$ and $P$. The phase
transition point $n_s^*$ where $H\to 0$ and $n_s^{\rm act}$ shows a
cusp, does not depend on $\epsilon$. For $n_s>n_s^*$, $n_s^{\rm act}$
has a discontinuous behavior as $\epsilon\to 0$. Indeed $n_s^{\rm act}$
increases linearly with $n_s$ for $\epsilon\le 0$ whereas it saturates
to a finite value for $\epsilon>0$.}
\label{Hnsact}
\end{figure}

The same behavior applies for $\epsilon>0$. For a fixed (reduced)
number $n_p$ of producers, $H$ decreases as the number of speculators
increases (see Fig. \ref{Hnsact}). As in the standard MG
\cite{CM,CMZe}, there is a phase transition separating a symmetric
phase ($H=0$), for $n_s>n_s^*(n_p)$, and an asymmetric phase ($H>0$).
The main difference is that in the symmetric phase many speculators
refrain from playing: The number of active speculators $n_s^{\rm act}$
saturates to a finite value as the number of speculators
increases. The behavior of $n_s^{\rm act}$ is characterized by a cusp
at $n_s^*$ (see Fig. \ref{Hnsact}) which becomes more pronounced as
$\epsilon\to 0^+$.

\begin{figure}
\centerline{\psfig{file=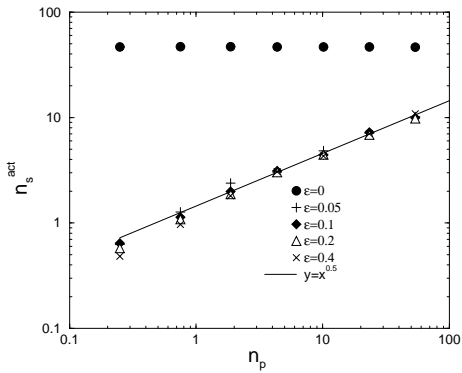,width=6cm}}
\caption{Number of active speculators $n_s^{\rm act}$ as 
a function of the number of producers $n_p$ for $P=4,~ n_s=62.5$
and several values of $\epsilon$.}
\label{nsnp}
\end{figure}

In the symmetric phase $n_s>n_s^*$, the market is kept {\em marginally
efficient}  in a dynamic way, that is, the market is
efficient in the long run (i.e. $H=0$), while locally in time, it may
be not efficient. Consequently, a fraction of speculators alternate
periods of activity, in which he trades, and inactivity, in which he
just watches the market, waiting for more favorable times. The level $n_s^{\rm act}$ of activity of
speculators is just barely sufficient to exploit the information injected
into the market by producers, thus making the market efficient. 
Fig. \ref{nsnp} shows that the number $n_s^{\rm act}$ of active 
speculators vanishes in the absence of producers and it increases 
as $n_s^{\rm act}\sim\sqrt{n_p}$ as the number of producers increases:
this is also the quantity of information put into the market by
the producers and asserts the validity of the proposed fundamental
grand-canonical mechanism.

The phase diagram  of the symmetric phase with respect to $n_s$, $n_p$
 and $\eps$ is quite complex, hence will be studied in details in a 
forthcoming publication, but let us discuss roughly its structure. At fixed $n_s$,

Again, the behavior for $\epsilon\le 0$ is quite different: The activity
does not vanish with $n_p$ but it rather remains finite even with 
$n_p=0$.

\subsection{Market crashes}

\begin{figure}
\centerline{\psfig{file=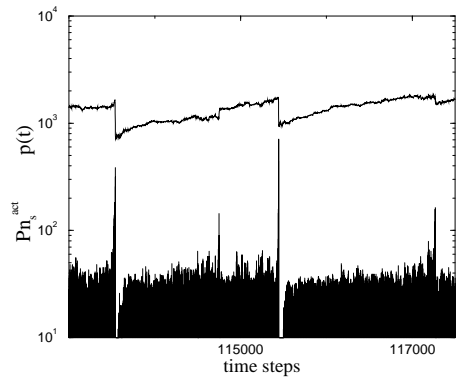,width=6cm}}
\caption{Snapshot of the dynamics for $n_s=300$, $n_p=10$ ($P=32$) and
$\epsilon=0.01$. The price $p(t)$ (top) and speculators contribution to the
volume $Pn_s^{\rm act}$ (bottom) is shown.}
\label{Crash}
\end{figure}

Finally we remark that fig. \ref{Hnsact} shows that the fluctuations
of $n_s^{\rm act}$ around its average diverge with $n_s$. This is due
to the fact that as $n_s$ increases the stationary state which agents
reach becomes more and more unstable. A snapshot of the dynamics deep
in the symmetric phase is reported in fig. \ref{Crash}. The market
repeatedly undergoes catastrophic events, i.e. crashes: Just after a
crash, all speculators refrain from trading and $n_s^{\rm act}\simeq
0$. This leaves the arbitrage opportunities created by producers
unexploited. Then speculators gradually gain confidence and start
trading again. The volume of speculation reaches a constant average
value, but fluctuations of $n_s^{\rm act}$ gradually increase. This
process reaches a point when a large number of speculators start
rushing into the market all at once. This causes violent shocks to
the market which causes discontinuities in the price,
i.e. crashes\footnote{Given the symmetry of the model, positive jumps
in $p(t)$ are as likely as negative ones.}.  The market crash drives
speculators away from the market and the dynamics starts anew. The
frequency of crashes increases with $n_s$ and it decreases with
$\epsilon$.

In spite of its simplicity, the model provides a quite realistic picture
of such a complex phenomenon as a market crash.

\section{Stylized Facts}

\subsection{Volatility and volume clustering}

Volatility clustering is a very well known stylized fact. It is known
that the volatility has algebraically decaying auto-correlation, and
accordingly that the returns activity is clustered in time, which is
an easy pattern to detect with naked eyes. 

\begin{figure}
\centerline{\psfig{file=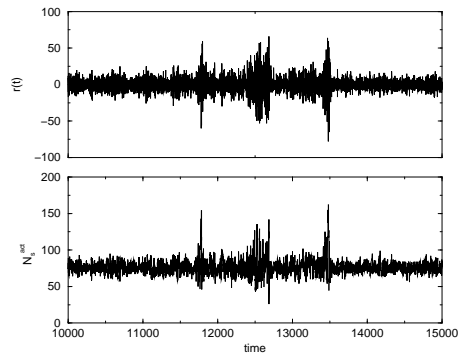,width=6cm}}
\caption{Return and number of speculators in the market  versus time. The volatility is clustered,
as is the volume ($P=16$, $S=2$, $N_s=501$, $N_p=1001$, $\eps=0.01$)}
\label{volclust}
\end{figure}

Fig. \ref{volclust} illustrates this phenomenon for the present model
(we define the volatility at time $t$ as $\sigma(t)=|A(t)|$).  The
deeper one goes in the symmetric phase -- i.e. the larger $n_s$ -- the
more high volatility regions appear clustered. Indeed market crashes
occur more and more frequently as $n_s$ increases. 

The volatility auto-correlation is known to be algebraically decaying,
typically as $\tau^{-0.3}$ in real financial markets \cite{Stanleyvol}.
Volatility auto-correlation is
known to be related to volume correlation \cite{Stanleyvol}.

\begin{figure}
\centerline{\psfig{file=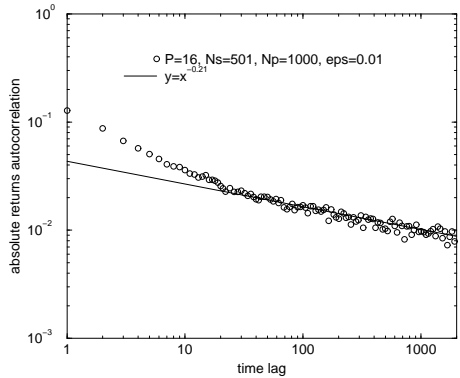,width=6cm}}
\caption{Autocorrelation of the absolute returns ($P=16$, $S=2$,
$N_s=501$, $N_p=1000$, $\eps=0.01$). The straight line as a $-0.21$ slope}
\label{autocorr}
\end{figure}

Figure \ref{autocorr} shows that the long ranged correlation of
volatility occurs also in the present model, with an exponent that can 
be close to that of real markets: it  depends quantitatively on the parameters
$n_s,~n_p$ and $\eps$ of the model, and exponents ranging from $0.09$ 
up to $0.6$ have already been observed, so that the let us only mention 
that this behavior is not universal, and can disappear if the parameters are  extremal.
 A systematic quantitative
study of this behavior in the whole phase space of the model, appears
quite demanding and will be pursued elsewhere.

Note furthermore that
the same behavior (with an exponent $-0.65$) was also found recently
in the MG with evolving capitals \cite{CCMZ}. This suggests that
volatility clustering is a generic feature of financial markets and
not an universal one, as also suggested recently in
ref. \cite{BouchMG}.

\subsection{Return and volume histograms}

\begin{figure}
\centerline{\psfig{file=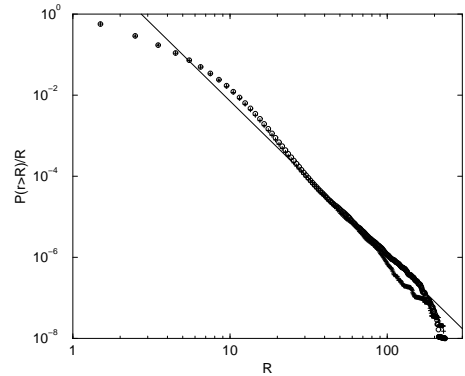,width=6cm}}
\caption{Cumulative function of the returns $R$ divided by the return $R$  (circles: positive returns, x: negative returns) ($P=16$, $S=2$,
$N_s=1001$, $N_p=1200$, $\eps=0.01$); the continuous line has a slope
of $-3.8$, close to the one observed in financial markets.}
\label{histret}
\end{figure}

In real market, the probability distribution function (pdf) of returns
is known to have fat tails with exponent -4 on average \cite{Stanleyret}. Figure
\ref{histret} shows that the MG presented here can reproduce fat tail
behavior. Note however that the value of the exponent depends on the
parameters $n_s$, $n_p$ and $\eps$. For instance, the exponent of the tails 
decreases as $n_s$ increases at fixed $n_p$, i.e. the tails become fatter 
and fatter as the number of prospective speculators increases while the 
gain opportunities remain fixed\footnote{We postpone the detailed
study of this point to a forthcoming publication}. The volume was also
found to have a power-law tail distribution. Figure \ref{histret} shows
typical returns histogram; parameter have been adjusted in order to obtain
 an exponent close to that of real markets \cite{Stanleyret}.

\section{Conclusion}

The modified MG presented here is able to reproduce qualitatively a
whole range of stylized facts. Most importantly, its quite rich
behavior can be studied analytically \cite{CMRZ} along the lines of
refs. \cite{CMZe,MCZ}.  At odds with the standard MG, speculators in
this model market are extremely sensitive to the number of producers,
and behave sensibly. 

All the features discussed in the model crucially depends on the fact
that agents neglect their market impact, i.e. behave as price takers.
As soon as agents start to account even approximately for their market
impact, as in refs. \cite{MCZ,MC}, the situation changes dramatically.
In particular: {\em i)} the phase transition disappears, {\em ii)} the
dynamics converges to one of exponentially many states where each
speculator either plays one and the same strategy at all times, or he
doesn't play at all, {\em iii)} volatility clustering and fat tailed
distribution of returns disappear. The latter means, in particular,
that all the stylized facts about financial markets crucially depends
on the fact that agents behave as price takers.

These conclusions -- strictly speaking-- only hold in the highly
simplified world described by our model. In real markets things are of
course much more complex: agents trade at different frequencies, over
different shares and with a large number of instruments such as
derivatives; their importance varies greatly, from big investment
funds and banks to small gamblers. Will the simplified picture, which
results from our simple model, survive even when all these issues have
been taken into account?

Partial positive answers have already been derived in recent work
where, for example, the existence of a phase transition has been shown
to persist when different types of agents are present \cite{MMM}, when
agents have different weights \cite{CCMZ} and when they trade at
different frequencies \cite{PM}. Here we showed that a phase
transition still persists when agents are allowed not to trade.  At
any rate, the present work opens a route to the understanding of how
complex phenomena -- such as volatility auto-correlation, fat tails in
the pdf of returns and market crashes -- may arise in financial
markets. This route may not be unique, but it is definitely worth of
investigation.

We acknowledge interesting discussions with Neil F. Johnson. This work has been
supported in part by  the Swiss National Funds for
Scientific Research.

\end{document}